\documentclass[11pt]{article}
\usepackage{amssymb}
\usepackage{amsmath}
\usepackage{color}
\usepackage{verbatim}
\usepackage[colorlinks,linkcolor=blue,citecolor=green]{hyperref}
\usepackage{graphicx}
\allowdisplaybreaks[3]

\title{Wave solutions for hyperbolic systems}
\author{Natale Manganaro$^1$,  Alessandra Rizzo$^2$\\
\\
 \small  $^1$ Department of Mathematical, Computer,\\
  \small Physical and Earth Sciences (MIFT)\\
  \small University of Messina, \\
  \small V.le F. Stagno D'Alcontres 31, 98166 Messina, Italy \\
\small e.mail: nmanganaro@unime.it \\
 \small  $^2$ Department of Mathematics and Computer Science,\\
  \small University of Palermo, \\
\small via Archirafi 34, 98123 Palermo, Italy \\
\small e.mail: alessandra.rizzo07@unipa.it}
\date{}
  
\begin{document}
\maketitle
\begin{abstract}
In this paper we propose a reduction procedure for determining generalized  travelling waves for first order quasilinear hyperbolic nonhomogeneous systems. The basic idea is to look for solutions of the governing model which satisfy a further set of differential constraints. Some applications are given for a barotropic fluid with a source term. 
%\textit{Key words:} term1 (phrase1), term2 (phrase2), .... .
\end{abstract}

{\bf Keywords:}  Differential Constraints. Travelling waves. Ideal fluid. 

{\bf MSC:}  35C07, 35C05, 35L40.

\section{Introduction}

Determining exact solutions of partial differential equations (PDEs) is of great interest not only from a theoretical point of view but also for possible applications. To this end, along the years, many mathematical approaches have been proposed most of them based on group  analysis (i. e. classical and non-classical symmetries \cite{bk, bc}, weak symmetries \cite{or}, conditional symmetries \cite{lw} (see also \cite{mel1})). 

Within such a theoretical framework, in 1964 J. J. Yanenko proposed the Method of Differential Equations \cite{jan} and he applied it to the fluid-dynamics equations. In order to explain better the basic idea of such an approach and also for further convenience, we give the following simple example.

Let us consider the PDE
\begin{equation}
u_t+a(u) u_x=f(u). \label{ee1}
\end{equation}
A particular class of exact solutions admitted by (\ref{ee1}) are the famous travelling waves, where
\begin{equation}
u=U(\sigma), \quad \quad \sigma=x- s \, t, \label{tv}
\end{equation}
with $s$ constant. In order to calculate the travelling waves of (\ref{ee1}) we have to substitute the ansatz (\ref{tv}) into (\ref{ee1}) and solving the resulting ordinary differential equation. The function (\ref{tv}) satisfies also the linear PDE
\begin{equation}
u_t + s u_x =0. \label{v1}
\end{equation}
Therefore, if we wont to determine the travelling waves of (\ref{ee1}), we can look for the particular solutions of (\ref{ee1}) which satisfy also (\ref{v1}). In such a case, since the equation (\ref{v1}) selects the class of exact solutions of (\ref{ee1}) we are looking for, they play the role of differential constraints. Of course, since the unknown $u(x,t)$ must satisfy (\ref{ee1}) along with (\ref{v1}) an overdetermined system is obtained and some compatibility conditions must be required. In this simple case, it is easy to verify that the equations (\ref{ee1}) and (\ref{v1}) are always compatible.  

More in general, given a system of PDEs
\begin{equation}
F^i(x,t, \mathbf{U},\mathbf{U}_x, \mathbf{U}_t, \mathbf{U}_{xx},\mathbf{U}_{xt},...)=0; \quad \quad i=1,..,N \label{ov1}
\end{equation}
Yanenko proposed to append to it a further set of differential constraints
\begin{equation}
G^k(x,t, \mathbf{U},\mathbf{U}_x, \mathbf{U}_t, \mathbf{U}_{xx},\mathbf{U}_{xt},...)=0; \quad  \quad k=1,..,M  \label{ov2}
\end{equation}
and to look for the exact solutions of (\ref{ov1}) and (\ref{ov2}). Of course, the compatibility of the overdetermined system (\ref{ov1}) and (\ref{ov2}) must be required.  The method is general and in fact it includes many of the known approaches for determing exact solutions of PDEs. Unfortunately such a generality leads to an huge an complicated alghorytm such that, without any further hypotesis, the method is not always useful for studying problems of interest for the applications.  To overcome such a difficulty, in \cite{fsy}-\cite{rsy} it was required that (\ref{ov1}) and (\ref{ov2}) are in involution (i. e. no new differential relations can be obtained from them by differentiation). The involutivness requirement simplifies the algorthyms of the method, in particular for hyperbolic systems of equations. In fact, many results concerning wave problems described by hyperbolic systems are obtained \cite{ms1}-\cite{mrv}. Moreover, an interesting application of the method to a parabolic model was given in \cite{rive}.

Within such a framework, here we develop a reduction procedure which permits to determine generalized travelling wave solutions for first order quasilinear nonhomogeneous hyperbolic systems. 

The paper is organized as follow. In section $2$ we recall  briefly the algorithm of the method of differential constraints applied to hyperbolic systems. In section $3$ we illustrate how the use of the $k-$Riemann invariants may simplifies such a procedure. In section $4$ an approach for characterizing generalized travelling waves is developed. Finally some conclusione and final remarks are given in section $5$.

\section{General procedure}
In this section we illustrate the procedure related to the Method of Differential Constraints for a first order quasilinear strictly hyperbolic system.  Let us consider the quasilinear system
\begin{equation}
\mathbf{U}_t + A\left( \mathbf{U} \right) \mathbf{U}_x=\mathbf{B}\left( \mathbf{U} \right) \label{hs}
\end{equation}
where $\mathbf{U} \in \mathbf{R}^N$ is the field vector, $A$ the $N \times N$ matrix coefficients, $\mathbf{B}  \in \mathbf{R}^N$ the source vector, while $t$ and $x$ denote, respectively, time and space coordinates. We assume the hyperbolicity (in the $t-$direction) of (\ref{hs}) and we denote with $\lambda^i \left( \mathbf{U} \right)$ the eigenvalues of $A$ (characteristic speeds) while the corresponding right and left eigenvectors are indicated, respectively, by $\mathbf{d}^i \left( \mathbf{U} \right)$ and $\mathbf{l}^i \left( \mathbf{U} \right)$. Moreover we assume $\lambda^i \neq \lambda^j, \; \; \forall i\neq j$ (i.e. system (\ref{hs}) is stricly hyperbolic). We choose $\mathbf{d}^i$ and $\mathbf{l}^i$ such that the orthonomal condition is satisfied $(\mathbf{d}^i \cdot \mathbf{l}^j= \delta^{ij})$. We add to (\ref{hs}) the set of differential constraints
\begin{equation}
\mathbf{C}^i(x,t,\mathbf{U}) \cdot \mathbf{U}_x=q^i(x,t,\mathbf{U}) \quad \quad i=1,...,M \leq N  \label{vi2}
\end{equation}
where the functions $C^i$ and $q^i$ are still not specified. For requiring the compatibility between (\ref{hs}) and (\ref{vi2}) we differentiate them with respect to $t$ and $x$, so that we find
\begin{eqnarray}
&&\left( \frac{\partial \mathbf{C}^i}{\partial t}+\frac{\partial \mathbf{C}^i}{\partial \mathbf{U}}\frac{\partial \mathbf{U}}{\partial t}\right) \cdot \mathbf{U}_x +\mathbf{C}^i \cdot \left( \frac{dB}{dx}-\frac{d (A U_x)}{dx} \right)=\frac{\partial q^i}{\partial t}+\frac{\partial q^i}{\partial \mathbf{U}}\frac{\partial \mathbf{U}}{\partial t}   \label{cv} \\
&& \nonumber \\
&&\left( \frac{\partial \mathbf{C}^i}{\partial x}+\frac{\partial \mathbf{C}^i}{\partial \mathbf{U}}\frac{\partial \mathbf{U}}{\partial x}\right) \cdot \mathbf{U}_x +\mathbf{C}^i \cdot \mathbf{U}_{xx}=\frac{\partial q^i}{\partial x}+\frac{\partial q^i}{\partial \mathbf{U}}\frac{\partial \mathbf{U}}{\partial x} \label{cv1}
\end{eqnarray}
where $\frac{d}{dx}$ means for total derivative with respect to $x$. In order to eliminate $\mathbf{U}_{xx}$ between (\ref{cv}) and (\ref{cv1}) it follows soon that the vectors $\mathbf{C}_i$ must belong to the subspace of the left eigenvectors of A so that, owing to  the strictly hyperbolicity of (\ref{hs}), without loss of generality we can choose $\mathbf{C}_i = \mathbf{l}^i$ and therefore we prove that the most general first order differential constraints admitted by (\ref{hs}) have the form
\begin{equation}
\mathbf{l}^i \cdot \mathbf{U}_x=q^i(x,t,\mathbf{U}) \quad \quad i=1,...,M \leq N  \label{vi4}
\end{equation}
The case of a great interest for nonlinear wave problem is when the number of the constraints is $M=N-1$. In this case, from (\ref{hs}) and (\ref{vi4}) we have
\begin{eqnarray}
&&\mathbf{U}_t=\mathbf{B}-\sum_{i=1}^{N-1} q^{i} \lambda^i \mathbf{d}^i - \pi  \lambda^N \mathbf{d}^N \label{ut} \\
&& \mathbf{U}_x=\sum_{i=1}^{N-1} q^i \mathbf{d}^i + \pi \mathbf{d}^N \label{ux}
\end{eqnarray}
where $\pi(x,t)$ is arbitrary. If we wont that the overdetermined system (\ref{hs}), (\ref{vi4}) is in involution, from (\ref{ut}), (\ref{ux}) we have to require that $\mathbf{U}_{tx}=\mathbf{U}_{xt} \; \forall \pi$, so that the following compatibility conditions are obtained
\begin{eqnarray}
&& q_t^i + \lambda^i q_x^i + \mathbf{\nabla}q^i \left( \mathbf{B} - \sum_{j=1}^{N-1} q^j \left( \lambda^j - \lambda^i \right) \mathbf{d}^j \right)+ \nonumber \\
&&+ \sum_{j=1}^{N-1} \sum_{k=1}^{N-1} q^j q^k \left( \lambda^j- \lambda^k \right)\mathbf{l}^i \mathbf{\nabla d}^j \mathbf{d}^k+ \nonumber \\
&& + \sum_{k=1}^{N-1} q^k \left( \mathbf{l}^i  \left(  \mathbf{\nabla d}^k \mathbf{B} -\mathbf{\nabla B} \, \,  \mathbf{d}^k  \right) +q^i \mathbf{\nabla} \lambda^i \mathbf{d}^k   \right)=0 \label{comp1} \\
&& \nonumber \\
&& \left( \lambda^i-\lambda^N \right) \mathbf{\nabla}q^i \mathbf{d}^N+ \sum_{k=1}^{N-1}q^k \left( \lambda^k-\lambda^N \right) \mathbf{l}^i \left( \mathbf{\nabla d}^k \mathbf{d}^N -\mathbf{\nabla d}^N \mathbf{d}^k \right) + \nonumber \\
&& +\mathbf{l}^i \left( \mathbf{\nabla d}^N \mathbf{B} -\mathbf{\nabla B} \, \, \mathbf{d}^N \right)+ q^i \mathbf{\nabla}\lambda^i \mathbf{d}^N=0 \label{comp2}
\end{eqnarray}
where $\nabla=\frac{\partial}{\partial \mathbf{U}}$ and $i=1...(N-1)$.

Furthermore, from (\ref{ut}) and (\ref{ux}) we obtain
\begin{equation}
\mathbf{U}_t+ \lambda^N \mathbf{U}_x=\mathbf{B}+\sum_{i=1}^{N-1} q^i \left( \lambda^N  - \lambda^i \right) \mathbf{d}^i \label{equa}
\end{equation}
Since the left-hand side of system (\ref{equa}) involves the derivate of $\mathbf{U}$ along the characteristics associated to $\lambda^{(N)}$, equations (\ref{equa}) can be integrated by using the standard method of characteristics. By substituting the resulting solutions into the constraints (\ref{vi4}) we get (see \cite{mel1} for more details)
\begin{equation}
\mathbf{l}^i (\mathbf{U}_0 (x)) \cdot \frac{d \mathbf{U}_{0} (x)}{dx}=q^i \left( x,0,\mathbf{U}_0 \right) \quad \quad i=1 \dots (N-1) \label{con5}
\end{equation}
where $\mathbf{U}_{0}(x)=\mathbf{U}(x,0)$. Since the $N$ initial conditions $\mathbf{U}_0 (x)$ must satisfy the $N-1$ constraints (\ref{con5}) the solutions which can be obtained by integration of (\ref{equa}) are determined in terms of one arbitrary functions.

It is of some interest to notice that in the case where $\mathbf{B}=0$ and $q^i =0$, the compatibility conditions (\ref{comp1}) and (\ref{comp2}) are identically satisfied and the procedure above illustrated permits to determine the classical simple wave solutions.

\section{An alternative approach}
The crucial point of the method of differential constraints is to study and possibly to solve the compatibility conditions (\ref{comp1}), (\ref{comp2}). Unfortunately it is a very hard task not  only to find the general solution of (\ref{comp1}), (\ref{comp2}) but also to determine particular solutions of such an overdetermined system. In order to simplify the analysis of (\ref{comp1}), (\ref{comp2}), quite recently in \cite{jmr1} an alternative approach based on the use of the Riemann invariants was proposed.

The basic idea is the following. We fix one of the characteristic speeds of (\ref{hs}) (for instance we can choose, without loss of generality, $\lambda^N$) and we compute its Riemann invariants defined by
\begin{equation}
\mathbf{\nabla}R^\alpha \cdot \mathbf{d}^N =0, \quad \alpha=1,..., N-1 \label{ri11}
\end{equation}
It is well known that associated to $\lambda^N$ there exist $N-1$ Riemann invariants whose gradients are linearly independent (see for instance \cite{smo}). Therefore, owing to (\ref{ri11}), we can write
\begin{equation}
\nabla R^\alpha = \sigma_{\beta}^{\alpha}\,  \mathbf{l}^\beta, \quad \alpha, \beta = 1,..., N-1 \label{nr}
\end{equation}
where $\sigma_{\beta}^{\alpha}$ are the components of $\nabla R^\alpha$ with respect the basis of the left eigenvectors. Moreover, here and what follows, the greek indices vary from $1$ to $N-1$. Taking (\ref{nr}) into account, the constraints (\ref{vi4}) assume the form
\begin{equation}
\frac{\partial R^\alpha}{\partial x}=\sigma_{\beta}^{\alpha} \, q^\beta \label{rvin1}
\end{equation}
and, in turn, equations (\ref{equa}) give
\begin{equation}
\frac{\partial R^\alpha}{\partial t}+ \lambda^N \frac{\partial R^\alpha}{\partial x}= \sigma_{\beta}^{\alpha} \, \mathbf{l}^\beta \cdot \mathbf{B}+ \left( \lambda^N - \lambda^\beta \right)\sigma_{\beta}^{\alpha} \, q^\beta \label{requ}
\end{equation}
We choose one of the field variables of $\mathbf{U}$ (say for instance $v=u_j$) and we add to the $N-1$ equations (\ref{requ}) the $j-$th equation arising from (\ref{equa})
\begin{equation}
\frac{\partial v}{\partial t}+ \lambda^N \frac{\partial v}{\partial x}=B_j + \left( \lambda^N - \lambda^\beta \right)q^\beta d_{j}^{\beta} \label{vequ}
\end{equation}
where $B_j$ and $d_{j}^{\alpha}$ denote, respectively, the $j-$th component of $\mathbf{B}$ and $\mathbf{d}^\alpha$. Of course we can always choose $v$ in such a way the variable  transformation
\begin{equation}
R^\alpha=R^\alpha (\mathbf{U}), \quad v=u_j \label{transf}
\end{equation}
is not singular. Therefore, the equations (\ref{equa}) transform to (\ref{requ}) and (\ref{vequ}) while the constraints (\ref{vi4}) take the form (\ref{rvin1}). Integration of (\ref{requ}), (\ref{vequ}) along with (\ref{rvin1}) gives, through the change of variables (\ref{transf}), exact solutions of (\ref{hs}), (\ref{vi4}). Furthermore, by requiring the  involutiveness of the overdetermined system (\ref{rvin1})-(\ref{vequ}), the following compatibility conditions are obtained:
\begin{eqnarray}
&&\left( \lambda^\beta - \lambda^N \right) \sigma_{\beta}^{\alpha} \, \frac{\partial q^\beta }{\partial v}=\left( \left( \lambda^N - \lambda^\beta \right) \frac{\partial \sigma_{\beta}^{\alpha}}{\partial v} - \sigma_{\beta}^{\alpha} \frac{\partial  \lambda^\beta }{\partial v} \right) q^\beta + \nonumber\\
&&+\frac{\partial}{\partial v} \left( \sigma_{\beta}^{\alpha} \, \mathbf{l}^\beta \cdot \mathbf{B} \right) \label{c1} \\
&& \frac{\partial w^\alpha}{\partial R^\gamma} \, z^\gamma - \frac{\partial z^\alpha}{\partial R^\gamma} \, w^\gamma +\frac{\partial w^\alpha}{\partial v}\left( B_j + \left( \lambda^N - \lambda ^\gamma \right) q^\gamma d_{j}^{\gamma}\right)=0 \label{c2}
\end{eqnarray}
where, for simplicity, we set
\begin{equation}
w^\alpha= \sigma_{\beta}^{\alpha} \, q^\beta, \quad \quad z^\alpha=\sigma_{\beta}^{\alpha} \, \mathbf{l}^\beta \cdot \mathbf{B}- \lambda^\beta \sigma_{\beta}^{\alpha} \, q^\beta. \label{wz}
\end{equation}
The analysis of the equations (\ref{c1}), (\ref{c2}) is not so hard as that of (\ref{comp1}), (\ref{comp2}). In fact, we notice that the $N-1$ equations (\ref{c1}) characterize a linear ODE-like system in the unknown $q^\alpha$ which, due the strictly hyperbolicity of (\ref{hs}), can be written in normal form. Once the functions $q^\alpha$ are determined from (\ref{c1}), substituting them in (\ref{c2}), we find a set of $N-1$ structural conditions that the coefficients of the system (\ref{hs}) must satisfy to guarantee the compatibility among (\ref{hs}) and (\ref{vi4}). 

Two significant cases where (\ref{c1}) and (\ref{c2}) are solved under suitable structural conditions are the following (see \cite{jmr1} for a more general analysis).

\vspace{0.2cm}
\noindent
{\it i)} We assume $q^\alpha =0$ so that from (\ref{c1}), (\ref{c2}) we find
\begin{equation}
\sigma_{\beta}^{\alpha} \, \mathbf{l}^\beta \cdot \mathbf{B}= F^\alpha (R^\gamma) \label{co1}
\end{equation}
where $F^\alpha$ are not specified functions. If the structural condition (\ref{co1}) is satisfied, then, from (\ref{rvin1}) we find $R^\alpha=R^\alpha (t)$ and taking (\ref{requ}), (\ref{vequ}) into account,  exact solutions of (\ref{hs}) are obtained by solving the system
\begin{eqnarray}
&&\frac{d R^\alpha}{d t}=F^\alpha (R^\gamma) \label{c1eq1}, \quad \quad  \\
&&\frac{\partial v}{\partial t} + \lambda^N\left( v, R^\alpha (t) \right) \frac{\partial v}{\partial x}=B_j \left( v, R^\alpha (t) \right). \label{c1eq2}
\end{eqnarray}
In passing we notice that the equations (\ref{c1eq1}) are decoupled from (\ref{c1eq2}). In fact, once $R^\alpha (t)$ are determined from (\ref{c1eq1}), exact solutions of the governing system can be obtained by solving the quasilinear non-autonomous PDE (\ref{c1eq2}). Furthermore, when $\mathbf{B}=0$ also $F^\alpha =0$, the compatibility conditions (\ref{co1}) are identically satisfied and the equations (\ref{c1eq1}), (\ref{c1eq2}) characterize the simple waves
\begin{eqnarray}
&&R^\alpha \left( \mathbf{U} \right)=k^\alpha  \nonumber \\
&& u_j=v=v_0 (\xi), \quad \quad x=\lambda^N \left( v_0 \left( \xi \right), k^\alpha \right)t +\xi \nonumber
\end{eqnarray}
where $k^\alpha$ are arbitrary constants and $v_0 (x)=v(x,0)$.

\vspace{0.2cm}
\noindent
{\it ii)} We require
\begin{equation}
\sigma_{\beta}^{\alpha} \left( \mathbf{l}^\beta \cdot \mathbf{B} - \lambda^\beta q^\beta \right)= F^\alpha (R^\gamma), \quad \quad \sigma_{\beta}^{\alpha}\, q^\beta = G^\alpha (R^\gamma). \label{cc2}
\end{equation}
so that, the condition (\ref{c1}) is identically satisfied, while from (\ref{c2}) we find
\begin{equation}
\frac{d G^\alpha}{d R^\beta}\,  F^\beta -\frac{d F^\alpha}{dR^\beta}\, G^\beta=0. \label{k1}
\end{equation}
where $F^\alpha$ and $G^\alpha$ are not specified. The functions $q^\alpha$ can be calculated by soving the linear algebraic system (\ref{cc2})$_2$ while from (\ref{cc2})$_1$ supplemented by (\ref{k1}) a set of $N-1$ structural conditions which must be satisfied by the coefficients of (\ref{hs}) in order that such a procedure holds, are obtained.

Furthermore, equations (\ref{requ}) assumes the form
\begin{equation}
\frac{\partial R^\alpha}{\partial t}+ \lambda^N \frac{\partial R^\alpha}{\partial x}= F^\alpha +\lambda^N G^\alpha \label{e31}
\end{equation}
while the constraints (\ref{vequ}) specialize to
\begin{equation}
\frac{\partial R^\alpha}{\partial x}=G^\alpha. \label{e32}
\end{equation}
It is of interest to notice that  if $\lambda^N (R^\gamma)$, then equations (\ref{e31}) are decoupled from (\ref{vequ}). Thus, once $R^\alpha (x,t)$ are determined from (\ref{e31}), exact solutions of (\ref{hs}) can be obtained by integrating the PDE (\ref{vequ}) by means of the method of caracteristics. Of course the corresponding initial data must obey the constraints (\ref{e32}).

\section{Generalized travelling waves}
Within the framework of the method of differential constraints, the main aim of this section is to develop a reduction procedure which permits to determine a class of exact solutions which generalize the classical travelling waves. 

To this end we append to system (\ref{hs}) the following constraints
\begin{equation}
\mathbf{U}_t+ s \mathbf{U}_x=\mathbf{F}\left( \mathbf{U} \right) \label{c01}
\end{equation}
where $s$ is a constant and $\mathbf{F}\left( \mathbf{U} \right)$ is unspecified. We decompose the vector $\mathbf{U}_x$ along the basis of the right eingenvectors
\begin{equation}
\mathbf{U}_x=\pi_j \mathbf{d}^j \label{de}
\end{equation}
so that, from (\ref{hs}) and (\ref{c01}) we have
\begin{equation}
\mathbf{U}_t=\mathbf{F}- s \pi_j \mathbf{d}^j, \quad \quad \mathbf{U}_x=\pi_j \mathbf{d}^j  \label{hh}
\end{equation}
where
\begin{equation}
\pi_i=\frac{1}{\lambda^i -s}\, \mathbf{l}^i \cdot \left( \mathbf{B}- \mathbf{F} \right). \label{pp}
\end{equation}
It is simple to verify that the compatibility between relations (\ref{hh}) leads to
\begin{equation}
F_i \frac{\partial \pi_s}{\partial u_i}=l_{k}^{s}\left( \frac{\partial F_k}{\partial u_i}d_{i}^{j}-\frac{\partial d_{k}^{j}}{\partial u_i} F_i\right)\pi_j \label{compa}
\end{equation}
Once $\pi_i$ are determined by solving the linear PDEs system (\ref{compa}), from (\ref{pp}) we find $\mathbf{F}$ and by integration of (\ref{hh}) a class of exact solutions of (\ref{hs}) are found. Such a solutions generalize the classical travelling waves because when $\mathbf{F}=0$, the compatibility conditions (\ref{compa}) are identically satisfied and from (\ref{hh})  travelling wave solutions are found.

\vspace{0.2cm}
\noindent
{\bf Remark.} Owing to (\ref{hh}) along with (\ref{pp}), if there exists a value $\mathbf{U}^\star$ of the field $\mathbf{U}$ such that $\lambda^i \left( \mathbf{U}^\star \right) =s$, a singularity in the travelling wave solution may appear depending if $\left( \mathbf{l}^i \cdot \left( \mathbf{B} - \mathbf{F} \right) \right)_{U=U^\star}$ vanishes or not. In the last case a sub-shock appears \cite{ruggeri}.

\vspace{0.2cm}
\noindent
As an example, here we apply our procedure the Euler system describing a barotropic fluid
\begin{eqnarray}
&&\rho_t +u \rho_x + \rho u_x=0 \label{e1} \\
&&u_t +u u_x+\frac{c^2}{\rho} \rho_x=f(\rho, u) \label{e2}
\end{eqnarray}
where $\rho$ is the mass density, $u$ the velocity, $c=\sqrt{p_\rho}$ the sound velocity with $p(\rho)$ the pressure while $f(\rho, u)$ denotes a force term. The characteristics velocities of (\ref{e1}), (\ref{e2}) are
\begin{equation}
\lambda^1 = u-c, \quad \quad \lambda^2=u+c   \label{ed}
\end{equation}
while the corresponding left and right eigenvectors are

\begin{eqnarray}
&&\mathbf{l}^1=\frac{1}{2} \left(
1 ,  -\frac{\rho}{c} \right), \quad \quad    \mathbf{l}^2=\frac{1}{2} \left(
1 , \frac{\rho}{c}
\right)\label{m1}  \vspace{0.1cm}\\
&& \mathbf{d}^1= \left(
1  ,  -\frac{c}{\rho}
\right)^T, \quad  \quad  \mathbf{d}^2= \left(
1 , \frac{c}{\rho}
\right)^T  \label{m2}
\end{eqnarray}
 where $T$ means for transposition.

Next, we consider the compatibility conditions (\ref{compa}) where, for simplicity, we assume $F_1 =0$. In such a case equations (\ref{compa}) assume the form
\begin{eqnarray}
&&F_2 \frac{\partial}{\partial u} \left( \pi_1 + \pi_2 \right) =0 \nonumber \\
&&F_2 \frac{\partial }{\partial u} \left( \pi_2 - \pi_1 \right)= \frac{\rho}{c}\left( \frac{\partial F_2}{\partial \rho} \left( \pi_1 + \pi_2 \right) +\frac{c}{\rho} \, \frac{\partial F_2}{\partial u} \left( \pi_2 - \pi_1 \right) \right) \nonumber
\end{eqnarray}
whose integration gives
\begin{equation}
\pi_2 - \pi_1 =\frac{1}{\Phi_u}\left( -\frac{\rho \alpha (\rho)}{c(\rho)} \Phi_\rho + \beta (\rho) \right), \quad \quad \pi_2 + \pi_1 = \alpha(\rho) \label{j1}
\end{equation}
where $\alpha(\rho)$ and $\beta(\rho)$ are not specified, while
$$
\Phi_u= \frac{1}{F_2 (\rho, u)}.
$$
Owing to (\ref{j1}), from (\ref{pp}) we determine $F_2$ as well as the source term $f(\rho, u)$. For instance, if $F_2 (u)$ we soon find
\begin{equation}
F_2=k_1 (u-s), \quad \quad f=k_1 (u-s) + k_1 \frac{c \beta}{\rho} \left( (u-s)^2 -c^2 \right) \label{fi}
\end{equation}
where $k_1$ is a constant. Furthermore the function $\alpha (\rho)$ must assume the form
$$
\alpha=-k_1 c \beta.
$$
In such a case, integration of (\ref{hh}) leads to
\begin{equation}
\rho=R(\sigma), \quad \quad u=s+\frac{a_0}{R(\sigma)}e^{k_1 t} \label{sol1}
\end{equation}
where $\sigma=x-st$, $a_0$ is a constant, while $R(\sigma)$ is given by solving the ODE
\begin{equation}
\frac{dR}{d \sigma}=-k_1 c(R) \beta (R). \label{sol2}
\end{equation}
We notice that the function $\beta (\rho)$ which is involved in the force term (\ref{fi})$_2$ is still not specified. For instance, if we choose
$$
\beta=\frac{\rho}{c}
$$
from (\ref{sol1}), (\ref{sol2}) the following generalized travelling wave solution is obtained
\begin{equation}
\rho=\rho_0 e^{-k_1 \left( x-st \right)}, \quad \quad u=s+\frac{a_0}{\rho_0} e^{k_1 \left( x- 2st \right)} \label{sol3}
\end{equation}
where $\rho_0$ is a constant. Finally we notice that the solution (\ref{sol1}), (\ref{sol2}) is obtained $\forall p(\rho)$.

\section{Conclusions and final remarks}
In this paper we developed a reduction procedure for determing a special class of exact solutions admitted by hyperbolic first order systems. The idea is to append to the original systems a further set of particular nonhomogeneous differential constraints which, in the homogeneous case, characterize travelling waves. Therefore the solutions we obtained generalizes the known travelling wave solutions.

To accomplish such a procedure, we are led to integrate the linear PDEs system (\ref{compa}) and, in turn, from (\ref{pp}) to calculate the vector $\mathbf{F}(\mathbf{U})$ involved in the constraints (\ref{c01}). Once $\pi_i$ or, equivalently, $\mathbf{F}$ are determined, generalized travelling waves of (\ref{hs}) can be obtained by solving the ODEs (\ref{hh}). We applied such a procedure to the hyperbolic system describing a barotropic fluid. The exact solutions we obtained for such equations are determined for any pressure law.

We conclude by noticing that the travelling waves are, in general, non admitted by non-autonomus systems. Our procedure can be applied also to systems like (\ref{hs}) where the matrix coefficients $A$ and/or the source vector $\mathbf{B}$ depend on the field $\mathbf{U}$ as well as on the variables $x$ and $t$. In fact, in such a case we have to require that the source vector $\mathbf{F}$ involved in the constraints (\ref{c01}) depends on $\left( \mathbf{U}, x, t \right)$. As a consequence, the compatibility conditions (\ref{compa}) must be modified by adding the derivatives of $\pi_i$ with respect to $x$ and $t$ and from (\ref{hh}) generalized travelling waves are obtained also in the non-autonomous case.\\
\subsection*{Acknowledgements} The authors thank the financial support of GNFM of the Istituto Nazionale di Alta Matematica. Furthermore N.M. thanks also the financial support of University of Messina through project FFABR UNIME 2023. A. R. thanks the financial support of the PRIN project MIUR Prin 2022, project code 1074 2022M9BKBC, Grant No. CUP B53D23009350006.

\end{document}